\documentclass[aps,pre,reprint,amsmath,amssymb,groupedaddress,floatfix]{revtex4-2}

\usepackage{graphicx}
\usepackage{placeins}

\newenvironment{publicationfigure}{\begin{figure}[htbp]\centering}{\end{figure}}
\newcommand{\publicationfigurewidth}{\linewidth}
\newcommand{\publicationdatafigureheight}{0.55\textheight}

\begin{document}

\title{Optimal staircase structure for bacterial adaptation in antibiotic landscapes}

\author{Kento Matsukawa}
\author{Shin-ichi Sasa}
\affiliation{Department of Physics, Kyoto University, Kyoto, Japan}
\author{Namiko Mitarai}
\affiliation{Niels Bohr Institute, University of Copenhagen, Copenhagen, Denmark}

\date{\today}

\begin{abstract}
Spatial heterogeneity in antibiotic environments drives the evolution of bacterial
resistance. However, how the discrete structure of these environments controls the time
required for adaptation of bacteria remains unclear.
We study this question using a stochastic model on a two-dimensional lattice whose axes
represent physical space and phenotype space, comparing different numbers of steps
while holding the overall spatial and concentration ranges fixed and scaling the migration
and mutation rates by their end-to-end first-passage times.
We find that increasing the number of steps accelerates the search for the next adapted
state but increases the cumulative cost of repeated population saturation; this competition
can produce an optimum at an intermediate number of steps.
Direct Gillespie simulations reproduce the predicted dependence of the adaptation time on
the number of steps, providing testable predictions for how discrete environmental
structure shapes the timescale of evolutionary adaptation.
\end{abstract}

\maketitle

\section{\label{sec:intro}Introduction}

Exposure to antibiotics drives the evolution of resistance through mutation and natural selection.
The emergence of antibiotic-resistant bacteria is a serious public health
problem~\cite{AMRCollaborators2022}, requiring a better understanding of the underlying mechanisms.
Recent experiments have shown that spatially heterogeneous antibiotic distributions can
qualitatively alter the emergence, spread, and maintenance of resistance~\cite{Zhang2011,Baym2016,Leale2018}.
In connected microfluidic environments, resistance emerged much faster under an
antibiotic gradient than in its absence, demonstrating that spatially structured drug
exposure can accelerate adaptation~\cite{Zhang2011}.
The MEGA-plate experiment, in which a concentration gradient was imposed on a large
agar plate, directly visualized how \textit{Escherichia coli} successively invades and
colonizes regions of higher antibiotic concentration, thereby showing how the spatial
organization of antibiotic exposure channels evolutionary adaptation~\cite{Baym2016}.

Several stochastic models have been proposed to study
the relationship between the spatial structure of the environment and evolution.
A stochastic source--sink model first described how spatial heterogeneity can facilitate
adaptation through mutation and migration~\cite{Hermsen2010}.
Building on this framework, the staircase model was introduced to describe how bacteria
acquire resistance step by step in a spatially discretized antibiotic gradient and was
analyzed both analytically and numerically~\cite{Hermsen2012}.
The model is defined on a two-dimensional lattice whose axes represent physical space and
phenotype space. Bacteria migrate through physical space and change phenotype through
mutation while their population grows.
That study showed that, in the presence of a concentration gradient, mutants can colonize new compartments
while avoiding competition with the wild type.
As a result, adaptation is markedly accelerated compared with a uniform environment.
It has also been reported that, depending on the structure of the mutational pathway
(a sequential increase of resistance or the crossing of a fitness valley), the effect
of a concentration gradient can be either to promote or to suppress
adaptation~\cite{Greulich2012}.
Furthermore, a previous study showed that bacterial motility can govern the speed of
adaptation by affecting the establishment and spread of resistant lineages~\cite{Piskovsky2023}.
Together, these studies have clarified how mutation, migration, and population dynamics
shape bacterial adaptation in spatially heterogeneous environments.

Here, we extend the staircase model by treating the number of environmental
and phenotypic steps $k$ as a control parameter.
Previous studies have primarily treated the number and size of the steps
as given~\cite{Hermsen2012,Hermsen2016}. By contrast, we compare systems with
different numbers of steps.
We vary $k$ while keeping the overall spatial extent, the antibiotic
concentration difference between the two ends, and the total carrying
capacity fixed.
Thus, increasing $k$ resolves the same spatial and concentration ranges into
a larger number of discrete steps.
To compare these systems on an equal footing, we introduce $k$-dependent
migration and mutation rates. These rates are chosen so that the mean
end-to-end first-passage times of a single bacterium in physical space and
phenotype space are independent of $k$.

We quantify the adaptation time by the mean first-arrival time (FAT) at the fully adapted state.
By decomposing the FAT into adaptation and population-saturation times, we derive an
approximate analytical expression for the FAT.
Increasing $k$ shortens the search time for the next adapted state at each step but
increases the cumulative cost of repeated population saturation.
Our analysis shows that, within the regime considered, there exists an optimal number
of steps $k^*$ that minimizes the FAT.
In an appropriate asymptotic regime, the optimum obeys the power law
$k^*+1 \sim (r^2/\mu v K_0)^{1/3}$, where $r, v, \mu$, and $K_0$ denote
the division rate, migration rate, mutation rate, and total carrying capacity, respectively.
Direct Gillespie simulations support the predicted dependence of the FAT on $k$,
including the location of the optimum.

The remainder of this paper is organized as follows.
Section~\ref{sec:model} defines the model, Sec.~\ref{sec:prelim} derives the analytical
FAT, Sec.~\ref{sec:results} analyzes the optimum and compares theory with simulations,
and Sec.~\ref{sec:concl} discusses the implications and limitations; technical
derivations are provided in the Appendices.

\section{\label{sec:model}Model and Setup}

We consider an extension of the staircase model~\cite{Hermsen2010,Hermsen2012}
in which the number of environmental and phenotypic steps $k$ is treated as
a control parameter.
As illustrated in Fig.~\ref{system_overview}, each bacterium occupies a site
$(i,p)$ in a two-dimensional state space, where $i$ denotes its physical patch
and $p$ its resistance phenotype.
The antibiotic concentration increases along the physical axis, and a
bacterium with phenotype $p$ can divide only in the region $i\leq p$.
The boundary of this region forms the staircase structure.
Migration and mutation correspond to transitions between neighboring sites
along the physical and phenotype axes, respectively, while bacteria undergo
division and death at each site.
We study the adaptation process from a population initially located at
$(1,1)$ until the fully adapted site $(k+1,k+1)$ is first reached.
\begin{publicationfigure}
    \includegraphics[width=\publicationfigurewidth]{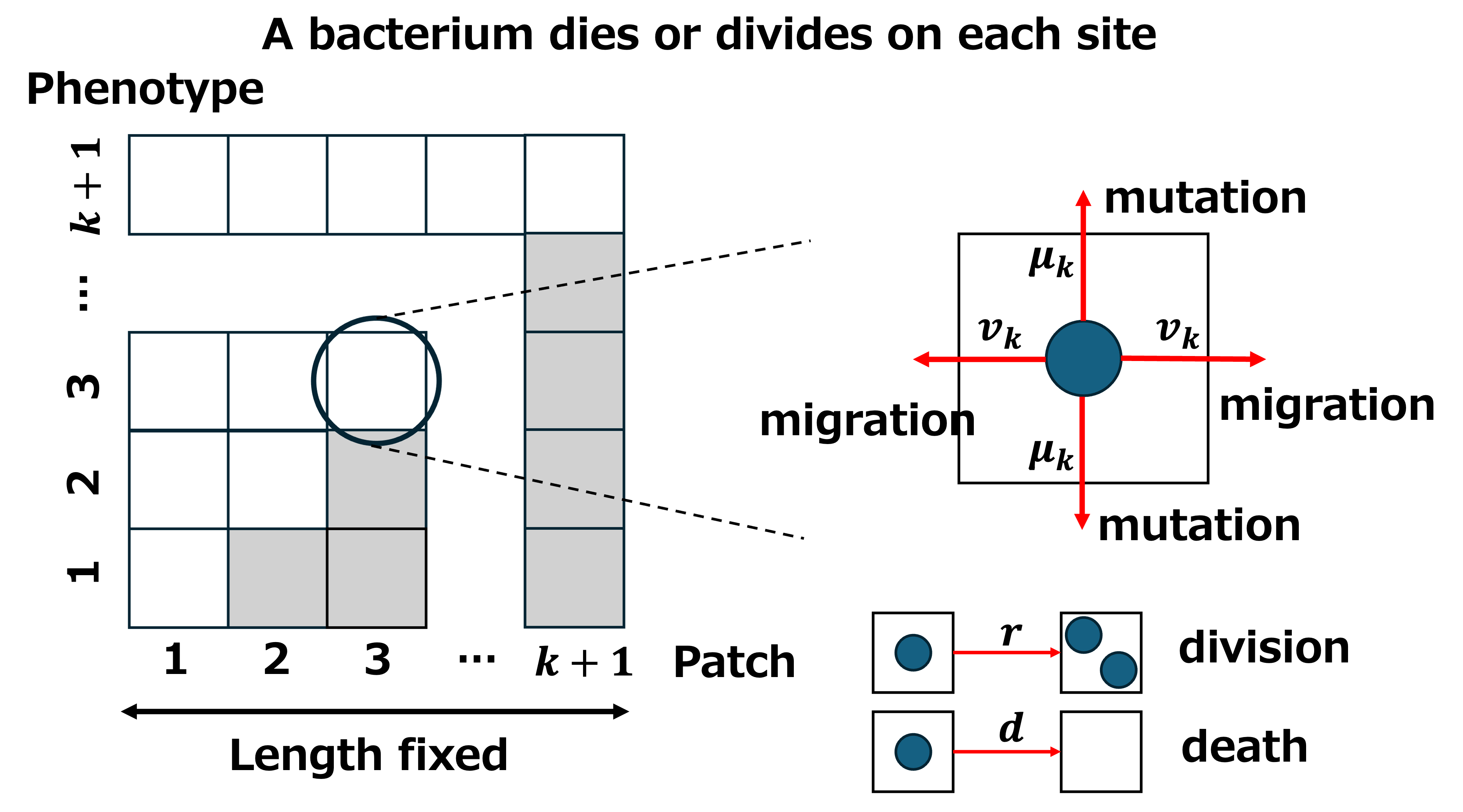}
    \caption{\label{system_overview}Overview of the system.
    The horizontal axis is the patch $i$ and the vertical axis is the phenotype $p$.
    Mutation corresponds to a transition along the phenotype axis
    $(p\rightarrow p \pm 1)$ and migration to a transition along the patch axis
    $(i\rightarrow i\pm 1)$; transitions across the boundaries do not occur.
    Division is possible in the region $i\leq p$.}
\end{publicationfigure}
\FloatBarrier
The physical space consists of $k+1$ patches arranged along a one-dimensional
domain and labeled by $i=1,\ldots,k+1$.
As $k$ varies, we keep the overall spatial extent and the antibiotic
concentration difference between the two ends fixed.
The antibiotic concentration increases monotonically from patch $1$ to patch
$k+1$, although the increments between neighboring patches need not be equal.
A bacterium in patch $i$ migrates to each neighboring patch at rate $v_k$, and
transitions beyond the boundaries are prohibited.
We choose $v_k$ so that the mean first-passage time of a single bacterium from
patch $1$ to patch $k+1$ is independent of $k$.
Taking $v=v_1$ as the migration rate in the reference system with $k=1$ gives
\begin{equation}
    v_k=\frac{k(k+1)}{2}v.
    \label{eq.rate_def_v}
\end{equation}
For large $k$, this relation has the same $k^2$ dependence obtained by keeping
the macroscopic diffusion coefficient fixed as the patch width decreases.
Details of the first-passage-time calculation are given in
Appendices~\ref{app:fpt} and~\ref{app:scaling_rates}.

The phenotype space consists of $k+1$ ordered resistance phenotypes labeled
by $p=1,\ldots,k+1$.
A bacterium with phenotype $p$ can divide in patches $i\leq p$, but cannot
divide in patches $i>p$.
Thus, a larger $p$ represents adaptation to a higher antibiotic
concentration.
When $k$ is varied, the least- and most-resistant endpoints are kept fixed,
while the number of intermediate phenotype levels changes.
Among the many possible genetic mutations, the model includes only those
that change the resistance phenotype.
Their net effect is represented by nearest-neighbor transitions
$p\rightarrow p\pm1$ at rate $\mu_k$.
Transitions beyond the two boundaries of phenotype space are prohibited.
As in physical space, $\mu_k$ is chosen so that the mean first-passage time
of a single bacterium from phenotype $1$ to phenotype $k+1$ is independent
of $k$.
Applying the same first-passage-time calculation gives
\begin{equation}
    \mu_k=\frac{k(k+1)}{2}\mu,
    \label{eq.rate_def_mu}
\end{equation}
where $\mu=\mu_1$ is the phenotype-transition rate in the reference system
with $k=1$.
Here, $\mu_k$ is an effective rate for transitions between neighboring
resistance phenotypes, rather than the total genetic mutation rate.

Figure~\ref{fig.setting} illustrates how the system changes with $k$ under
these scaling relations.
\begin{publicationfigure}
    \includegraphics[width=\publicationfigurewidth]{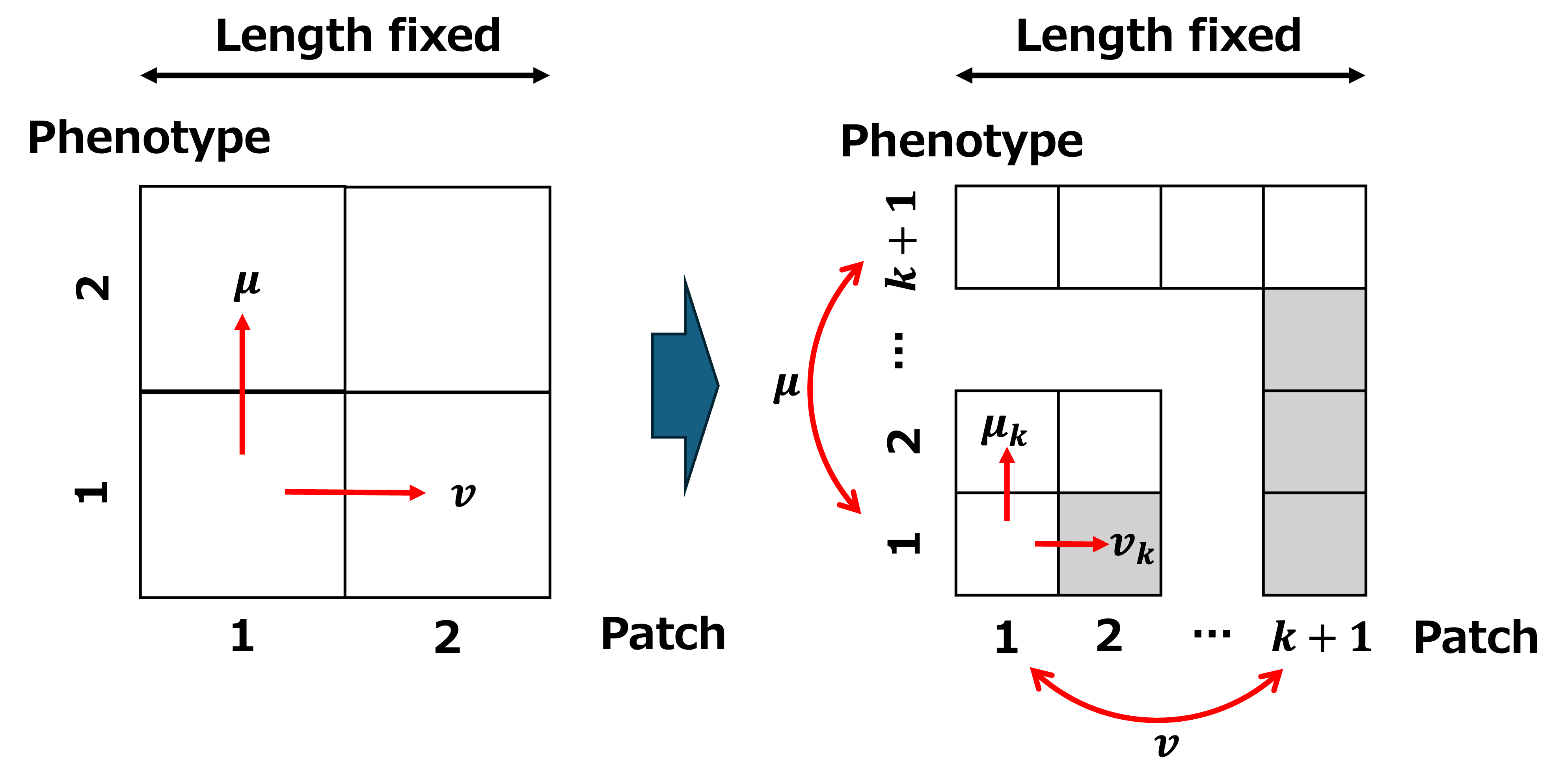}
    \caption{\label{fig.setting}Comparison of the reference system with $k=1$ and a staircase with $k$ environmental and phenotypic steps.
    The left panel shows the reference system with $k=1$.
    The right panel illustrates the corresponding system for a larger $k$, with the
    end-to-end migration and mutation first-passage times held fixed.}
\end{publicationfigure}
\FloatBarrier

In addition to migration and mutation, the population dynamics are governed
by density-dependent division and constant death.
To compare systems with different $k$ under the same total carrying capacity,
we keep $K_0$ fixed and distribute it equally among the $k+1$ patches.
The carrying capacity of each patch is therefore
\begin{equation}\label{eq.K}
    K=\frac{K_0}{k+1}.
\end{equation}
Writing $N_i$ for the total number of bacteria in patch $i$, summed over all
phenotypes, the per-capita division rate at site $(i,p)$ is
\begin{equation}
    r_{i,p} =
    \begin{cases}
        r\left(1-\dfrac{N_i}{K}\right), & i\leq p,\\
        0, & i>p,
    \end{cases}
\end{equation}
where $r$ is the intrinsic division rate in the low-density limit.
Thus, bacteria with different phenotypes in the same patch compete for a
common carrying capacity.
Bacteria die at a constant per-capita rate $d$ at every site, independent
of their physical patch and phenotype.

The analytical treatment is restricted to a parameter regime in which a
locally adapted population can persist and phenotypic transitions are rare
compared with migration.
Specifically, for every value of $k$ considered, we assume
$r>d+2v_k$ and $d>v_k\gg\mu_k$.
For an adapted population in an interior patch, division balances death and
outward migration at the local saturated population size
\begin{equation}\label{eq.N}
    N=\left(1-\frac{d+2v_k}{r}\right)K.
\end{equation}
The condition $r>d+2v_k$ therefore ensures that this population size is
positive.
Because the migration rate increases as the physical domain is divided into
narrower patches, the condition $d>v_k$ restricts the range of $k$ considered
in the analysis.
The separation $v_k\gg\mu_k$ represents the rarity of transitions between
resistance phenotypes relative to migration.
Since $v_k$ and $\mu_k$ have the same dependence on $k$, this separation is
equivalent to $v\gg\mu$ and is independent of $k$.

The rate parameters differ in how directly they can be related to
experimentally measured quantities.
For \textit{E.~coli} K-12 MG1655 under rich-medium growth conditions, the intrinsic
division rate $r$ is of the order of $2\,\mathrm{h}^{-1}$~\cite{Sezonov2007}, whereas
the death rate $d$ depends strongly on the environment~\cite{Regoes2004}.
The reference migration rate $v$ can be related to measurements of bacterial
diffusion.
For wild-type \textit{E.~coli}, the measured effective diffusion coefficient
is approximately
$200\,\mu\mathrm{m}^2/\mathrm{s}$~\cite{Kurzthaler2024}.
Converting this diffusion coefficient into the migration rate $v$ requires
the patch width and boundary conditions to be specified.
Likewise, $\mu$ is an effective transition rate between neighboring
resistance phenotypes rather than the total genetic mutation rate measured
experimentally~\cite{Lee2012}.

We quantify the speed of adaptation by the mean first-arrival time.
Starting from a population saturated at site $(1,1)$, we define
$\mathrm{FAT}(k)$ as the mean time until site $(k+1,k+1)$ is occupied for
the first time.
Our central questions are whether $\mathrm{FAT}(k)$ is minimized at an
intermediate number of steps and, if so, how the minimizing value $k^*$
depends on the model parameters.

\section{\label{sec:prelim}Preliminaries}

In this section, we introduce the assumptions used in the analytical
treatment and derive an approximate expression for $\mathrm{FAT}(k)$.
Simulations of the staircase model and previous studies suggest that the dominant adaptation pathway follows the sequence of sites
$(1,1)\rightarrow(2,2)\rightarrow\cdots\rightarrow(k+1,k+1)$~\cite{Hermsen2010,Hermsen2012}.
The analysis assumes that the population approaches local saturation at each site
$(i,i)$ before adaptation to the next site occurs. The transition from $(i,i)$ to
$(i+1,i+1)$ is referred to as climbing one step of the staircase and is decomposed into
two stochastic times.
The saturation time $\hat{T}_{\mathrm{sat}}^{(i)}$ is the first-passage time from the first occupation of site $(i,i)$ 
until its population reaches the local saturated population size $N$.
The adaptation time $\hat{T}_{\mathrm{adapt}}^{(i)}$ is the first-passage time from this locally saturated state 
until site $(i+1,i+1)$ is first occupied.

The sequential-adaptation assumption is motivated by the rarity of mutation compared with the other processes.
In the strong-selection weak-mutation regime, a mutant becomes established before the next mutation 
occurs~\cite{Gillespie1983,Gillespie1984}.
Such sequential adaptation has also been observed in microbial evolution experiments~\cite{Atwood1951}.
When the mutation supply is large, multiple advantageous mutant lineages can instead coexist and compete.
This regime, known as clonal interference~\cite{Sniegowski2010,Desai2007}, is not considered here.
Under these assumptions, the mean time $\mathrm{FAT}(k)$ for a bacterial population
saturated at $(1, 1)$ to reach $(k+1, k+1)$ can be written as
\begin{align}\label{TimeScaleSep}
    \mathrm{FAT}(k) = \left\langle\sum_{i=1}^{k}\hat{T}_{\text{adapt}}^{(i)} + \sum_{i=2}^{k}\hat{T}_{\text{sat}}^{(i)}\right\rangle,
\end{align}
where $\langle \cdot\rangle$ denotes the expectation value.
We set $T_{\text{adapt}}:=\langle\hat{T}_{\text{adapt}}^{(i)}\rangle$ and
$T_{\text{sat}}:=\langle\hat{T}_{\text{sat}}^{(i)}\rangle$.
Because each staircase step is treated as the same stochastic process, the
distributions of the adaptation time and the saturation time do not depend
on $i$.
Consequently, neither do $T_{\text{adapt}}$ and $T_{\text{sat}}$.
Therefore
\begin{align}
    \mathrm{FAT}(k) = kT_{\text{adapt}} + (k-1)T_{\text{sat}}.
\end{align}
The second sum starts at $i=2$ because the population is already saturated at $(1,1)$
in the initial state.
Below we obtain explicit expressions for $T_{\text{adapt}}$ and $T_{\text{sat}}$.

\subsection{\label{sec:tadapt}Derivation of $T_{\text{adapt}}$}

The derivation follows the method of Ref.~\cite{Hermsen2010}, with the transition from
$(i,i)$ to $(i+1,i+1)$ treated as a generic local adaptation step.
At an interior site, migration to two neighboring patches produces a total
migration loss $2v_k$.
At the boundary site $(1,1)$, the exact migration loss is instead $v_k$
because this site has only one neighboring patch.
To describe every step by the same local process, we approximate the
migration loss at all sites by the interior value $2v_k$.
Under this approximation, the saturated population at site $(i,i)$ fluctuates
around the value $N$ defined in Eq.~\eqref{eq.N}.
For analytical simplicity, we neglect these fluctuations and treat the
population size as the constant $N$.
Let $P(n,t)$ denote the probability that there are $n$ bacteria at site $(i,i+1)$ at time $t$,
with the target site $(i+1,i+1)$ not yet reached by time $t$.
The master equation satisfied by $P(n,t)$ is
\begin{align}\label{eq.master_eq}
    \frac{\partial P(n,t)}{\partial t}
    &= (\delta-v_k)(n+1)P(n+1,t) \nonumber\\
    &\quad +[r'(n-1)+\mu_k(N-n+1)]P(n-1,t) \nonumber\\
    &\quad -[(\delta+r')n+\mu_k(N-n)]P(n,t),
\end{align}
where $\delta=d+2v_k+\mu_k$ and $r'=d+2v_k$.
The target site $(i+1,i+1)$ is treated as an absorbing boundary for this first-passage problem.
We define the survival probability as $S(t)=\sum_n P(n,t)$.
Because the target is absorbing, $S(t)$ is the probability that site
$(i+1,i+1)$ has not been reached by time $t$.
Let $F(t)$ denote the probability density of the adaptation time
$\hat{T}_{\mathrm{adapt}}^{(i)}$.
It follows from the definition of $S(t)$ that $F(t)=-dS(t)/dt$.
To obtain $S(t)$, we consider the generating function $G(z,t)=\sum_n z^n P(n,t)$.
By deriving the partial differential equation satisfied by $G(z,t)$ from
Eq.~\eqref{eq.master_eq} and solving it, we obtain
\begin{align}\label{FATPDF}
    S(t) &= \left(\frac{be^{\mu_kt}}{b\cosh(bt) + \mu_k\sinh(bt)}\right)^a,\\
    F(t) &= -\frac{dS(t)}{dt}
    = \frac{\mu_k Nv_k \tanh(b t)}{b + \mu_k\tanh(bt)}S(t),
\end{align}
where
\begin{align}\label{eq.a_def}
    a=\mu_k N/(r'-\mu_k),
\end{align}
and $b=\sqrt{\mu_k^2 + v_k(r'-\mu_k)}$.

Since $T_{\text{adapt}} = \int_{0}^{\infty} t F(t) dt$, we have
\begin{align}\label{Tadapt_int}
    T_{\text{adapt}} = \int_{0}^{\infty} t\left(-\frac{dS(t)}{dt}\right)dt = \int_{0}^{\infty}S(t)dt.
\end{align}
Evaluating Eq.~\eqref{Tadapt_int} directly is difficult, but it can be computed
approximately by Laplace's method.
Writing $S(t) = e^{-af(t)}$, the exponent $f(t)$ has a minimum at the boundary $t=0$ of
the range of integration, with $f(0)=f'(0)=0$ and $f''(0)=b^2-\mu_k^2>0$.
For $a \gg 1$, which typically corresponds to $K_0 \gg 1$, the integral is therefore
dominated by the neighborhood of the boundary minimum at $t=0$.
Expanding $f(t)$ around $t=0$ and applying Laplace's method gives
\begin{align}\label{eq.adapt}
    T_{\text{adapt}} = \int_{0}^{\infty}S(t)dt
    &\simeq \int_{0}^{\infty}\exp\left[-\frac{a}{2}(b^2-\mu_k^2)t^2\right]dt,\nonumber\\
    &= \sqrt{\frac{\pi}{2a(b^2-\mu_k^2)}} = \sqrt{\frac{\pi}{2\mu_kv_k N}}
\end{align}
with details of the calculation given in Appendix~\ref{app:adapt}.

\subsection{\label{sec:tsat}Derivation of $T_{\text{sat}}$}
Once adaptation has occurred, two main situations can be distinguished.
In the first, the adapted bacterium dies or migrates to another site before it divides,
with probability $p\simeq (d+2v_k)/(r+d+2v_k)$, and adaptation has to take place
again.
The expected time $T_{\text{readapt}}$ until this readaptation can be computed in the
same way as Eq.~\eqref{Tadapt_int}.
When an adapted individual is lost without establishing itself, we approximate its loss
time by the mean adaptation time $T_{\text{adapt}}$.
We then treat the unadapted population state at that time as a new initial ensemble whose
total survival probability is normalized to unity.
The quantity $T_{\text{readapt}}$ is the arrival time of trajectories that adapt after
$T_{\text{adapt}}$, measured from the original time origin.
In this study we consider only a single readaptation and neglect the terms of order
$\mathcal{O}(p^2)$ and higher that originate from repeated failures, writing the
readaptation time as
\begin{align}\label{Treadapt_int}
    T_{\text{readapt}} = \int_{T_{\text{adapt}}}^{\infty} tF(t) dt.
\end{align}
In the limit of large $K_0$, the readaptation time is approximated by
\begin{align}\label{T_readapt}
    T_{\text{readapt}}  \simeq C T_{\text{adapt}},
\end{align}
where $C=1+e^{-\pi/4} - \int_{0}^{1}e^{-\pi u^2/4}du$ is a constant obtained in the
limit $K_0\rightarrow \infty$.
The details of the calculation are given in Appendix~\ref{app:adapt}.
In the second situation the bacterium divides and the population grows until it
saturates.
This saturation is required on every trajectory, including those on which the first
adapted individual was lost and adaptation had to take place again, so that to
$\mathcal{O}(p)$ we have
$T_{\text{sat}} \simeq (1-p)T'_{\text{sat}} + p(T_{\text{readapt}} + T'_{\text{sat}})
= T'_{\text{sat}} + pT_{\text{readapt}}$.
The mean growth time $T'_{\text{sat}}$ is obtained from the first-passage time of the
population size, following the method of Ref.~\cite{Sekimoto2022}.
In the limit of large $K_0$ this mean time $T'_{\text{sat}}$ becomes
\begin{align}\label{T'_sat}
    T'_{\text{sat}} = \frac{1}{r}\left(\ln\frac{NK}{K-N} + \gamma\right),
\end{align}
where $\gamma$ is the Euler--Mascheroni constant and the limit
$K_0\rightarrow\infty$ has been taken.
From the above,
\begin{align}\label{eq.sat}
    T_{\text{sat}} \simeq \frac{d + 2v_k}{r + d + 2v_k}C T_{\text{adapt}} + \frac{1}{r}\left(\ln\frac{NK}{K-N} + \gamma\right).
\end{align}
The details of the calculation are given in Appendix~\ref{app:sat}.

From Eqs.~\eqref{TimeScaleSep}, \eqref{eq.adapt}, and \eqref{eq.sat},
\begin{align}\label{eq.FATfunc}
    \mathrm{FAT}(k)
    &= k\sqrt{\frac{\pi}{2\mu_kv_kN}} \nonumber\\
    &\quad +C\frac{(k-1)(d+2v_k)}{r+d+2v_k}
      \sqrt{\frac{\pi}{2\mu_kv_kN}} \nonumber\\
    &\quad +\frac{k-1}{r}\left(\ln\frac{NK}{K-N}+\gamma\right).
\end{align}
The first term is the initial search time, that is, the time until the first individual
that mutates and then migrates appears out of the saturated population.
The second term is the cost of a failed establishment: the time until such an individual
appears again because the previous one was lost before dividing.
The third term is the cost of population saturation, which is incurred at every step on
every trajectory, whether the first adapted individual established itself immediately or
only after a failure.
The last term is larger by a factor $1+(d+2v_k)/r$ than it would be if saturation were
counted only on the trajectories that succeed at the first attempt; the difference is of
relative order $d/r$ and therefore affects only subleading terms.

Equation~\eqref{eq.FATfunc} reveals a competition between the search time for the next adapted individual 
and the repeated cost of population saturation.
Increasing $k$ raises both $\mu_k$ and $v_k$, thereby shortening the search time for the next adapted individual.
By contrast, increasing $k$ also increases the number of saturation events required to reach the fully adapted state.
The search-time contribution decreases roughly as $k^{-1/2}$, whereas the saturation contribution increases roughly as $k\ln(K_0/k)$.
Consequently, $\mathrm{FAT}(k)$ can attain a minimum at an intermediate value of $k$ in an appropriate parameter regime.

\section{\label{sec:results}Results}

In this section, we derive an asymptotic criterion for an interior minimum of
$\mathrm{FAT}(k)$ and
the value of $k$ that minimizes it.
The analytical predictions are then compared with numerical calculations and direct
stochastic simulations.
Before starting the analysis, we explicitly specify the range of $k$.
As mentioned in Sec.~\ref{sec:model}, we assume $r > d + 2v_k$ and
$d > v_k \gg \mu_k$.
Since $d > v_k = k(k+1)v/2$ must hold, the admissible $k$ satisfy $k(k+1) < 2d/v$.
For the values of $k$ of interest, the condition $k(k+1)<2d/v$ is well approximated by $k<\sqrt{2d/v}$ so that
we use $1 \leq k < \sqrt{2d/v} =: k_{\text{max}}$ throughout.

The physical model is defined for integer $k$.
To derive the scaling analytically, however, we temporarily regard $k$ as a real
variable and differentiate with respect to it.
The numerical minimizations of Sec.~\ref{sec:sim_scaling} likewise treat $k$ as
real, whereas the Gillespie simulations and the heat map of
Sec.~\ref{sec:speedup} use integer $k$.

\subsection{\label{sec:mincond}Asymptotic criterion for an interior optimum}

Equation~\eqref{eq.FATfunc} was first evaluated numerically to examine whether
$\mathrm{FAT}(k)$ has a minimum
at an intermediate value of $k$ over a wide range of parameters.
We generated $10^5$ parameter sets by sampling $K_0$, $v$, and $\mu$ 
from log-uniform distributions over $K_0\in(10^5,10^{10})$, $v\in(10^{-8},10^{-3})$, and $\mu\in(10^{-10},10^{-5})$, respectively.
The death rate $d$ was sampled uniformly from $(0.01,1)$, while $r=1$ was fixed.
We retained only the parameter sets satisfying $r>9d$, $d>99v$, $v>99\mu$, and $k_{\max}>1$, leaving 4574 sets.
For each parameter set, we evaluated Eq.~\eqref{eq.FATfunc} at $10^3$ equally spaced values of $k$ in $[1,k_{\max}]$ 
and counted the sign changes of the discrete derivative.
Among the 4574 parameter sets, 3256 had one extremum and 1318 had no extremum.
No parameter set exhibited more than one extremum.
Motivated by this numerical observation, we assume that $\mathrm{FAT}(k)$ has at most one extremum in the admissible range of $k$.
Under this empirically supported single-extremum assumption, a criterion for the minimum to occur at
$k>1$ is $\mathrm{FAT}(1)>\mathrm{FAT}(2)$.
Substituting $k=1$ and $k=2$ into Eq.~\eqref{eq.FATfunc} and retaining the leading terms for $K_0\gg1$, we obtain
\begin{align}
    \mathrm{FAT}(1) = \sqrt{\frac{\pi}{\mu v K_0}},\qquad
    \mathrm{FAT}(2) = \sqrt{\frac{2\pi}{3\mu v K_0}} + \frac{1}{r}\ln K_0.
\end{align}
The resulting asymptotic criterion can therefore be written as
\begin{align}\label{cond_min}
    \left(\frac{\mu v K_0}{r^2}\right)(\ln K_0)^2 < \pi\left(1-\sqrt{\frac{2}{3}}\right)^2.
\end{align}
The correction to the saturation bookkeeping discussed below Eq.~\eqref{eq.sat}
affects only subleading terms of order $d/r$ and therefore does not modify
Eq.~\eqref{cond_min}, nor the scaling of $k^*$ derived in the next section.

\subsection{\label{sec:kmin}Optimal number of steps}

Having established an asymptotic criterion for an interior minimum, we now determine the value $k^*$ that minimizes $\mathrm{FAT}(k)$.
Let $k^*$ be the value of $k$ satisfying
$\left(\partial \mathrm{FAT}/\partial k\right)_{k=k^*}=0$.
In what follows we consider two regimes of $k^{*}$.

For $v(k^*)^2\sim d$, that is $k^*\gg 1$, differentiating $\mathrm{FAT}(k)$ directly and
using this approximation gives
\begin{align}
    \sqrt{\frac{2\pi r^2}{\mu v K_0}}\left({k^*}^{-3/2} + \frac{5}{2}\frac{v}{r}{k^*}^{1/2}\right) = 2 - \ln\frac{rK_0}{v{k^*}^3}.
\end{align}
The left-hand side is positive whereas the right-hand side is negative because $K_0$
is assumed to be sufficiently large.
There is therefore no solution, and $\mathrm{FAT}(k)$ has no minimum.

Next, for $v(k^*)^2\ll d$, that is $k^* \ll k_{\text{max}}$, the equation satisfied by
$k^*$ is
\begin{align}
    -\frac{1}{2}\sqrt{\frac{2\pi}{\mu v K_0}}\frac{1}{x^{3/2}} + \frac{1}{r}\ln\frac{K_0 r}{dx} + \frac{2}{rx} = \frac{1}{r},
\end{align}
where
\begin{align}
    x = k^* + 1.
\end{align}
The stationarity equation determines the scaling that an interior solution must obey.
Using the implicit function theorem we can compute the logarithmic derivative
$\alpha_p = \partial \ln x / \partial \ln p$ of $x$ with respect to each parameter
$p\;(p\in \{r, \mu, v, K_0, d\})$.
These logarithmic derivatives determine the scaling exponents, yielding
Eq.~\eqref{eq.scaling}.
The conditions under which it holds are given in detail in
Appendix~\ref{app:scaling_kstar}.
\begin{align}
    k^{*} + 1 = A\left(\frac{r^2}{\mu v K_0}\right)^{1/3}.
    \label{eq.scaling}
\end{align}
Here $A$ is a numerical constant.

\subsection{\label{sec:speedup}Comparison of the adaptation speed}

We investigate how much the optimal staircase accelerates adaptation relative to the single-step case $k=1$.
This analysis is based on the analytical expression \eqref{eq.FATfunc} rather than on direct Gillespie simulations, 
because a sweep over this parameter range would be computationally prohibitive.
The analytical description is compared with Gillespie simulations in Sec.~\ref{sec:sim_theory}.
Figure~\ref{heatmap} shows the speedup ratio $\mathrm{FAT}(1)/\mathrm{FAT}(k^*)$, where $k^*$ is obtained 
by numerically minimizing Eq.~\eqref{eq.FATfunc}.
The optimal staircase can substantially reduce the adaptation time over a broad parameter range,
although the region in which a large speedup is obtained becomes smaller as $K_0$ increases.

\begin{publicationfigure}
    \includegraphics[width=\linewidth]{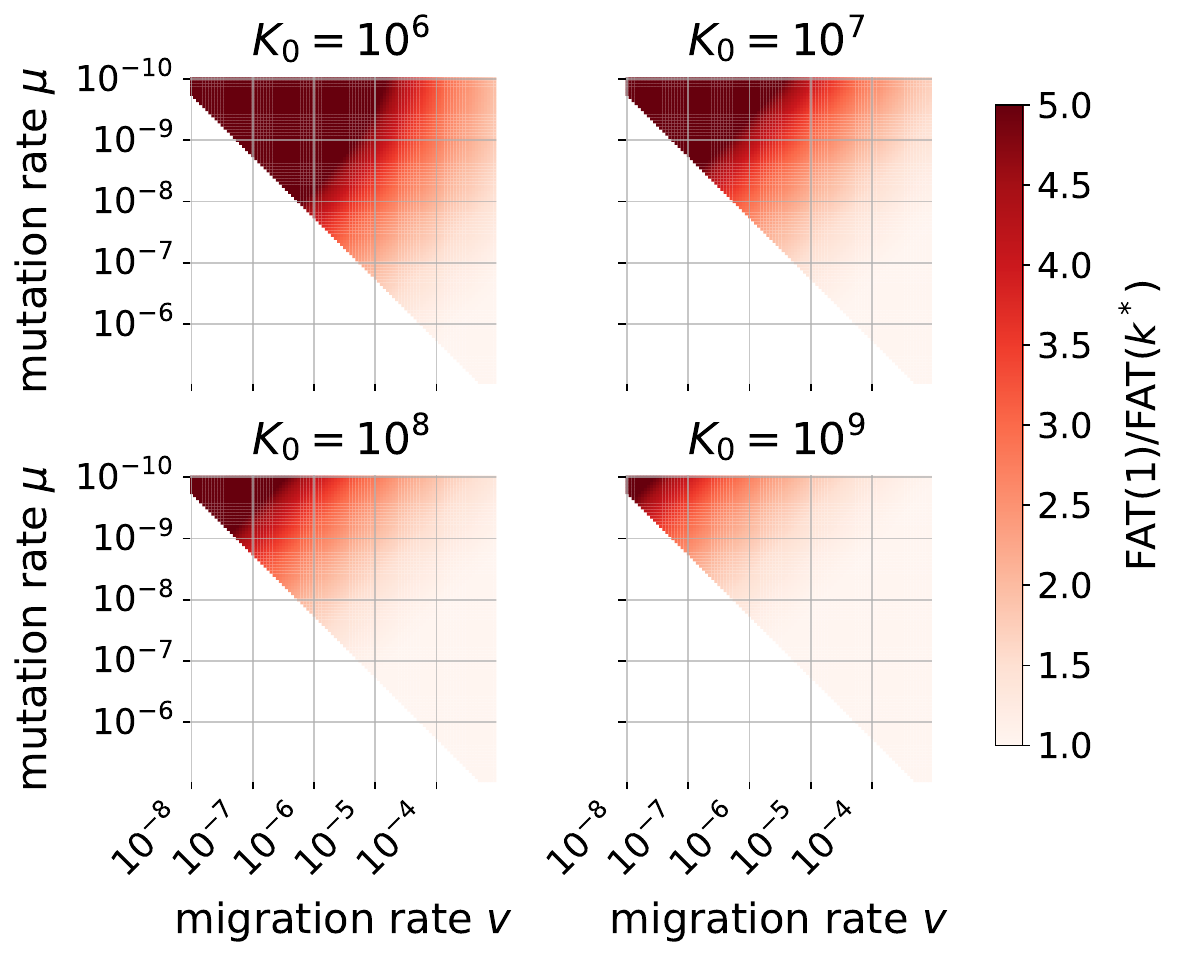}
    \caption{\label{heatmap}Speedup ratio $\mathrm{FAT}(1)/\mathrm{FAT}(k^*)$ obtained by numerical minimization of the analytical expression \eqref{eq.FATfunc}.
    The horizontal axis is $v$ and the vertical axis is $\mu$.
    The four panels correspond to $K_0=10^6,10^7,10^8$, and $10^9$, with $r=1$ and $d=0.1$ held fixed.
    The axes span $v\in(10^{-8},10^{-3})$ and $\mu\in(10^{-10},10^{-5})$ on a $100\times100$ logarithmically spaced grid.
    The white region corresponds to $v<50\mu$ and is excluded because it does not satisfy the theoretical assumption $v\gg\mu$.
    Values of $\mathrm{FAT}(1)/\mathrm{FAT}(k^*)$ larger than $5$ are clipped at the upper limit of the color scale for visualization; the value $5$ has no particular physical significance.
    For each grid point, $k^*$ is determined by evaluating Eq.~\eqref{eq.FATfunc} at successive integer $k$ from $k=1$ up to $\mathrm{round}(k_{\text{max}})-1$ and taking the first minimum.}
\end{publicationfigure}

\subsection{\label{sec:sim}Simulations}

\subsubsection{\label{sec:sim_theory}Comparison with stochastic simulations}

The analytical description is tested by direct numerical simulations using the Gillespie algorithm.
Each simulation resolves all $k+1$ patches simultaneously.
The rates $\mu_k$, $v_k$, $d$, and $r_{i,p}$ are those defined in Sec.~\ref{sec:model}, 
with the per-patch carrying capacity $K=\lfloor K_0/(k+1)\rfloor$.
An individual migrates to each existing neighboring patch at rate $v_k$.
The migration loss is therefore $2v_k$ per individual in the interior and $v_k$ at the two boundaries.
Thus, the boundary migration dynamics is treated exactly in the simulations.
The initial population size is nevertheless chosen using the same interior-site approximation
$N$ in Eq.~\eqref{eq.N} as in the analytical calculation.
Mutation is symmetric, with forward and backward rates both equal to $\mu_k$.
All phenotypes have the same intrinsic division rate $r$.
Each trial starts from $\lfloor(1-(d+2v_k)/r)K\rfloor$ individuals at $(1,1)$.
The trial terminates when site $(k+1,k+1)$ is occupied for the first time.
Every trajectory reached the target for all six parameter sets and all values of $k$, 
so no conditioning on successful trajectories was required.
For each value of $k$, we performed 300 independent Gillespie trials.

Figure~\ref{fig.sample_pattern} compares the Gillespie simulations with two theoretical evaluations of $\mathrm{FAT}(k)$.
The purple circles show the mean values of the Gillespie simulations, with error bars representing the standard error of the mean.
The blue solid curve is obtained by numerically evaluating Eqs.~\eqref{Tadapt_int} and \eqref{Treadapt_int}
without the Laplace approximation used to derive Eq.~\eqref{eq.FATfunc}.
The orange dashed curve shows the approximate analytical expression \eqref{eq.FATfunc}.
The Gillespie simulations are performed for integer $k$ up to $10$, 
whereas the theoretical curves are evaluated up to $\min(\lfloor k_{\max}\rfloor,50)$.
The gray dashed vertical line indicates this upper limit of the theoretical curves.
The triangular markers indicate the minima of the corresponding data or curves.

The Gillespie simulations and the numerically evaluated theory show the same qualitative behavior over the parameter ranges examined.
In particular, the theory captures the minimum at an intermediate $k$ in
Figs.~\ref{fig.sample_pattern}(a)--\ref{fig.sample_pattern}(c) and the nearly
monotonic increase of $\mathrm{FAT}(k)$ from $k=1$ in
Figs.~\ref{fig.sample_pattern}(d)--\ref{fig.sample_pattern}(f).
This agreement indicates that the sequential-adaptation assumption 
and the decomposition into adaptation and saturation times capture the leading behavior of the FAT.
The approximate expression \eqref{eq.FATfunc} systematically gives larger FAT values than the Gillespie simulations, 
with the ratio ranging from approximately $1.0$ to $1.5$ in the cases shown.

We next examine the accuracy of the Laplace approximation used to obtain Eq.~\eqref{eq.FATfunc}.
The blue and orange curves differ only in the evaluation of the adaptation and readaptation times, 
while the saturation contribution is identical.
The parameter $a=\mu_kN/(r'-\mu_k)$, defined in Eq.~\eqref{eq.a_def}, controls the accuracy of the Laplace approximation, 
which assumes $a\gg1$.
At $k=1$, the numerical and approximate evaluations of the adaptation time differ by $5.4\%$ in Figs.~\ref{fig.sample_pattern}(a)--\ref{fig.sample_pattern}(c), where $a\simeq4.5$,
and by $0.6\%$ in Figs.~\ref{fig.sample_pattern}(d)--\ref{fig.sample_pattern}(f), where $a\simeq45$.
The discrepancy decreases approximately as $1/a$ as $k$ increases.
The difference in the total FAT is smaller because the common saturation contribution becomes increasingly important at larger $k$.
The two theoretical curves give the same location of the minimum in every panel.
Thus, for sufficiently large $a$, the Laplace approximation captures the numerically
evaluated theoretical expression and gives the same location of the minimum in the
cases examined.

We finally compare how the model parameters affect the shape of $\mathrm{FAT}(k)$.
Comparing Figs.~\ref{fig.sample_pattern}(a) and \ref{fig.sample_pattern}(b),
increasing $v$ from $10^{-5}$ to $10^{-4}$ decreases the FAT and shifts its
minimum toward smaller $k$.
Comparing Figs.~\ref{fig.sample_pattern}(c) and \ref{fig.sample_pattern}(d),
increasing $K_0$ from $10^7$ to $10^8$ increases $a$ at $k=1$ from
approximately $4.5$ to $45$.
Correspondingly, the difference between the numerical and approximate theoretical curves decreases from $5.4\%$ to $0.6\%$.
Comparing Figs.~\ref{fig.sample_pattern}(e) and \ref{fig.sample_pattern}(f),
$K_0$ and $\mu$ are varied in opposite directions so that $\mu K_0$ remains
of the same order, with $v=10^{-3}$ fixed.
The resulting FAT curves have similar shapes, consistent with the leading dependence of the adaptation time on $\sqrt{\mu vK_0}$.
They do not coincide exactly because the saturation time also depends logarithmically on $K_0$.

\begin{publicationfigure}
    \includegraphics[width=\linewidth,height=\publicationdatafigureheight,keepaspectratio]{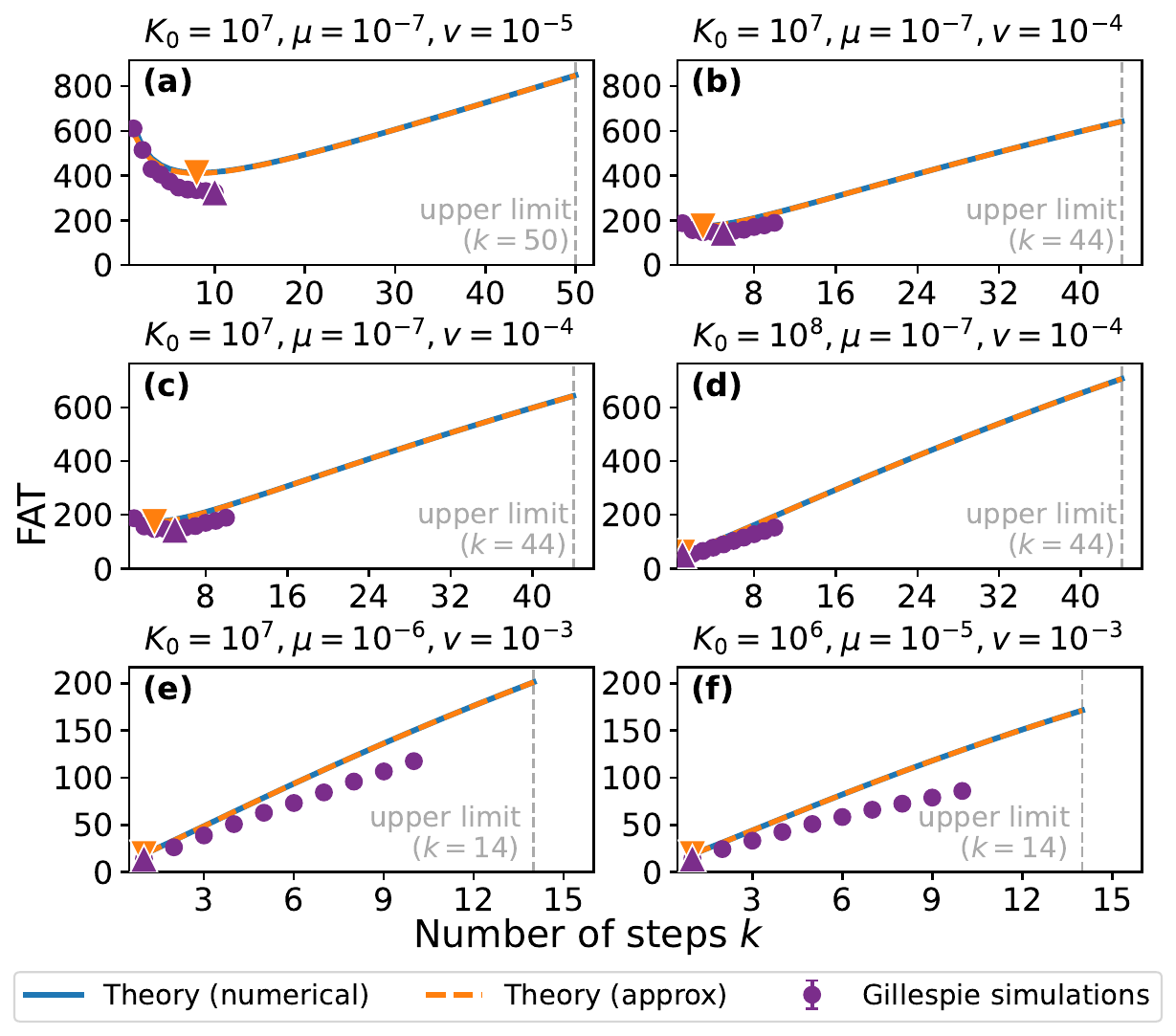}
    \caption{\label{fig.sample_pattern}Mean first-arrival time as a function of the number of steps $k$, obtained from theory and Gillespie simulations.
    The purple circles are the mean values over $n=300$ independent Gillespie trials,
    and the error bars show the standard error of the mean.
    Most error bars are smaller than the plotting symbols and are therefore not visible at this scale.
    The blue solid curve is the result of computing Eq.~\eqref{eq.FATfunc} with
    Eqs.~\eqref{Tadapt_int} and \eqref{Treadapt_int} evaluated by numerical
    integration, and the orange dashed curve is a plot of Eq.~\eqref{eq.FATfunc}; the
    two are distinguished by line style as well as by color.
    The gray dashed vertical line, labeled ``upper limit'' in the figure, marks the
    largest $k$ at which the theoretical curves are evaluated,
    $\min(\lfloor k_{\text{max}}\rfloor, 50)$ with
    $k_{\text{max}}=\sqrt{2d/v}$ the upper bound of the model.
    The triangular markers indicate the locations of the minima, the colors
    corresponding to the respective points or curves.
    The parameters used are $r=1.0, d=0.1$ in all panels, and
    (a) $K_0=10^7, \mu=10^{-7}, v=10^{-5}$,
    (b) $K_0=10^7, \mu=10^{-7}, v=10^{-4}$,
    (c) $K_0=10^7, \mu=10^{-7}, v=10^{-4}$,
    (d) $K_0=10^8, \mu=10^{-7}, v=10^{-4}$,
    (e) $K_0=10^7, \mu=10^{-6}, v=10^{-3}$, and
    (f) $K_0=10^6, \mu=10^{-5}, v=10^{-3}$.}
\end{publicationfigure}

\subsubsection{\label{sec:sim_scaling}Scaling of the optimal number of steps}
Having compared the analytical description with Gillespie simulations in the
computationally accessible regime, we now use the analytical expression to explore a
much broader parameter range.
Specifically, Eq.~\eqref{eq.FATfunc} was minimized numerically with respect to $k$,
with $k$ treated as a real variable on $[1, k_{\text{max}}]$.
The resulting $k^*+1$ was then plotted against $r^2/(\mu v K_0)$.
\begin{publicationfigure}
    \centering
    \includegraphics[width=\linewidth]{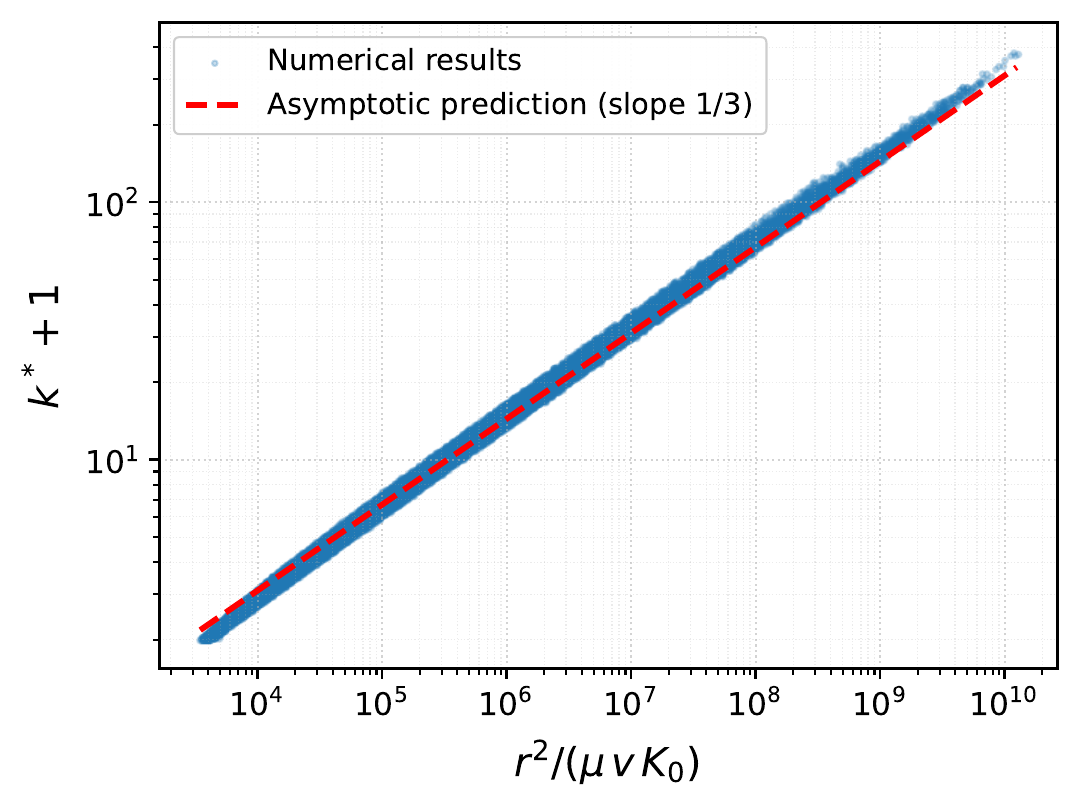}
    \caption{\label{scaling_kstar}Scaling of the optimum obtained by numerical
    minimization of the analytical expression \eqref{eq.FATfunc}.
    The horizontal axis is $r^2 / \mu v K_0$ and the vertical axis is $k^* +1$.
    The blue points are the $59{,}971$ values of $k^*+1$ obtained by minimizing
    Eq.~\eqref{eq.FATfunc} with respect to $k$, treated as a real variable on
    $[1, k_{\text{max}}]$.
    The parameter sets were generated by log-uniform sampling of $r \in (0.1, 5)$,
    $\mu \in (10^{-10}, 10^{-5})$, $v \in (10^{-8}, 10^{-3})$,
    $K_0 \in (10^{9}, 10^{10})$, and $d \in (0.01, 0.1)$
    ($3 \times 10^5$ raw samples), retaining those satisfying $r \geq 10d$,
    $d \geq 10v$, and $v \geq 50\mu$.
    A set was kept only if the minimum was interior, $k^* < 0.9\,k_{\text{max}}$, and
    satisfied $\mathrm{FAT}(k^*) < \mathrm{FAT}(1)$, leaving $59{,}971$ points.
    The red dashed line is the asymptotic prediction
    $\ln(k^*+1) = \frac{1}{3}\ln (r^2/\mu v K_0) + \text{Const}$, whose slope is
    fixed at $1/3$ and whose intercept $\text{Const}$ is obtained by fitting the data.}
\end{publicationfigure}
Figure~\ref{scaling_kstar} shows that the numerically obtained optima lie on or near the
asymptotic prediction with slope $1/3$, in agreement with the asymptotic analysis.
We stress that this comparison is between the asymptotic scaling and the numerical
minimization of Eq.~\eqref{eq.FATfunc}; it does not involve Gillespie simulations.
The parameter sets were not filtered by $a$.
Evaluating $a = \mu_k N/(r'-\mu_k)$ at $k^*$ for the retained sets gives a median of
$3.8\times10^{2}$.
Moreover, $98.5\%$ of the retained sets satisfy $a>10$.
Thus, the condition $a\gg1$ assumed in deriving Eq.~\eqref{eq.adapt} holds for the great
majority of the retained parameter sets.

\section{\label{sec:concl}Concluding Remarks}

This work extends the staircase model by treating the number of steps $k$ as a control
parameter while holding the system size fixed. The resulting approximate analytical
expression for the adaptation time, $\mathrm{FAT}(k)$, is given by
Eq.~\eqref{eq.FATfunc}.
$\mathrm{FAT}(k)$ can have a minimum at $k\neq 1$. Together with the numerical
observation that $\mathrm{FAT}(k)$ has at most one extremum, Eq.~\eqref{eq.FATfunc}
gives the asymptotic criterion in Eq.~\eqref{cond_min} for such a minimum.
The analytical calculation also showed that the value $k=k^*$ that minimizes the FAT
obeys the scaling given by Eq.~\eqref{eq.scaling}.
The results also show that the optimal staircase can substantially reduce the adaptation
time relative to the single-step case.

\subsection{\label{sec:interp}Interpretations of the results}

Both $r$ and $\sqrt{\mu v K_0}$ have dimensions of inverse time.
We refer to $r$ as the intrinsic growth rate and to $\sqrt{\mu v K_0}$ as the
characteristic adaptation rate.
In terms of these, Eq.~\eqref{eq.scaling} can be rewritten as
$k^*+1 \sim \left(r/\sqrt{\mu v K_0}\right)^{2/3}$ and Eq.~\eqref{cond_min} as
$\left(\sqrt{\mu v K_0}\,\ln K_0/r\right)^2 < \pi (1-\sqrt{2/3})^2$.
The optimal number of steps is thus controlled by the ratio of the intrinsic growth
rate to the characteristic adaptation rate.
The characteristic adaptation rate can be regarded as the rate at which the bacterial
population as a whole explores the two-dimensional space.
The optimum therefore reflects a trade-off between searching for the adapted state and
growing to fill a patch.
When $\sqrt{\mu v K_0}$ is small, the process of searching for the next adapted state
is rate-limiting.
Increasing the number of steps $k$ then increases $\mu_k$ and $v_k$, so that adaptation
per step becomes faster and a larger $k$ is favorable.
When $\sqrt{\mu v K_0}$ is large, on the other hand, the cost of saturating the
population at each step becomes dominant, and the $k$ that minimizes the time required
for adaptation becomes smaller.

In the characteristic adaptation rate, $\mu$ and $v$ appear in the form of a geometric
mean.
The geometric mean arises from the combination of the sequential mutation--migration process and the population effect.
Because adaptation requires mutation followed by migration, the short-time cumulative probability for a single bacterium to complete one adaptation step grows as $\mu v t^2/2$.
For a population of $N$ bacteria, the probability that no bacterium has completed this process therefore decays approximately as $\exp(-N\mu v t^2/2)$.
Consequently, the characteristic first-arrival time scales as $1/\sqrt{N\mu v}$, giving rise to the geometric mean of $\mu$ and $v$.
The details are discussed in Appendix~\ref{app:geomean}.

\subsection{\label{sec:limits}Limitations and future work}

For the sake of the analytical calculation, we assumed in this paper that the growth
rate is the same for all phenotypes.
Mutations that confer drug resistance, however, can be accompanied by a reduction of
the growth rate in the absence of the drug, that is, by a fitness
cost~\cite{Sandegren2008,Andersson2010}.
If fitness is to be taken into account in order to obtain a more realistic
description, one possible approach is to let $s$ be the fitness of the state $(k+1, k+1)$ relative to
the state $(1,1)$ and to interpolate between them by a geometric sequence.
We conjecture that the qualitative behavior, such as the appearance of a minimum, would
remain unchanged in this case, although we have not verified this.
Analytically, $r, \mu, v$ would presumably be replaced by effective rates; whether the
structure of the scaling survives such a replacement remains to be determined.
It has also been reported that the fitness cost associated with resistance can be
alleviated by compensatory mutations~\cite{Schrag1997}.
Incorporating such compensatory evolution, in addition to a phenotype-dependent growth
rate, is left for future work.

The model treats the mutation and migration rates as symmetric, but in reality
mutations in the reverse direction are often rare and the migration rates are often
asymmetric as well.
We conjecture that the qualitative behavior would be unchanged in such cases and that a
similar scaling may hold once the rates are replaced by effective rates, but it remains
to be determined whether this is so.

A further limitation is that each patch is treated as well mixed, so all individuals in
a patch are implicitly able to contribute to migration into the next patch.
In a spatially extended experimental system such as the MEGA-plate experiment, however,
access to the next drug region is largely restricted to lineages near the exposed
advancing front, while beneficial mutants arising behind the front can be spatially
blocked by cells ahead of them~\cite{Baym2016}.
The effective mutation supply for further advance may therefore be set by the front
population rather than by the total population within a patch.

Experimental tests of the model require estimates of the parameters $r$, $\mu$, $v$,
$K_0$, and $d$.
Among these parameters, the mutation rate $\mu$ is particularly difficult to measure
directly.
The model, however, provides an analytical expression for the adaptation time at $k=1$.
If the remaining parameters are independently estimated, comparison with an
experimentally measured adaptation time could provide an indirect estimate of $\mu$.
This estimate could then be used to test the existence and scaling of an optimal number
of steps.

\begin{acknowledgments}
    This work was supported by JSPS KAKENHI (Grant Nos.~JP25K00923 and JP26H00383).
    N.M. acknowledges Novo Nordisk Foundation support (NNF21OC0068775).
\end{acknowledgments}

\appendix

\section{\label{app:fpt}First Passage Time for Markov Processes}

This appendix briefly reviews the first-passage time in Markov chain processes.
Let $P_{\alpha}(t)$ be the probability that a system is in the state $\alpha\in S$ at
time $t$, where $S$ is the set of all states.
The probability $P_{\alpha}(t)$ is assumed to satisfy the master equation
\begin{align}\label{Master_Eq}
    \frac{d}{dt}P_{\alpha}(t) = \sum_{\beta}M_{\alpha \beta}P_{\beta}(t).
\end{align}
Here $M_{\alpha \beta}$ is assumed not to depend on time $t$.
It is also assumed to satisfy the probability-conservation condition
$\sum_{\alpha}M_{\alpha \beta} =0$.

To characterize the first-passage time, we divide the state space $S$ into the set $G$
of goal states and its complement $G^c$, that is,
$S = G \cup G^c$.
Starting from an initial state $i \in G^c$, let $s(t)$ denote the state of the system at
time $t$.
The first-passage time $\tau$ is the first time at which $s(\tau)\in G$.

Because $\tau$ is determined only by the trajectory up to the first arrival at $G$, its
distribution is unaffected by how the motion is modified after that arrival.
The dynamics can therefore be modified so that the system remains in $G$ after first
entering it. Removing the transitions $G \rightarrow G^c$ gives
\begin{align}\label{Modified_Dynamics}
    \frac{d}{dt}P^*_{\alpha}(t) = \sum_{\beta} M^*_{\alpha\beta}P^*_{\beta}(t).
\end{align}
Here $\alpha, \beta \in G^c$, and because there is an outflow into $G$, we do not
necessarily have $\sum_{\alpha} M^*_{\alpha\beta} = 0$.
Thus, Eq.~\eqref{Modified_Dynamics} describes a system in which probability is not
conserved within $G^c$.

Under the modified dynamics a trajectory that has once reached $G$ does not return to
$G^c$.
Therefore $P^*_{G^c}(t) = \sum_{\alpha \in G^c}P^*_{\alpha}(t)$ is precisely the
probability that the goal has not been reached by time $t$; it starts from
$P^*_{G^c}(0) = 1$ and decreases monotonically.
Since this decrease is the probability of reaching the goal during $[t, t+dt]$,
\begin{align}
    \mathrm{prob}\bigl(\tau \in [t, t+dt]\bigr)
    &= P^*_{G^c}(t) - P^*_{G^c}(t+dt)\nonumber \\
    &\simeq - \frac{dP^*_{G^c}(t)}{dt}dt,
\end{align}
so that $-dP^*_{G^c}/dt$ is the probability density function of the first-passage time
$\tau$.

For the initial state $\lvert P^*(0)\rangle = \lvert i\rangle$, the formal solution of
Eq.~\eqref{Modified_Dynamics} is
\begin{align}
    \lvert P^*(t)\rangle = e^{M^* t}\lvert i\rangle,
\end{align}
where $\lvert P^*(t)\rangle$ denotes the vector with components
$P^*_{\alpha}(t)$ for $\alpha\in G^c$.
Using $\langle 1 \rvert$ we write
$P^*_{G^c}(t) = \langle 1 \vert P^*(t)\rangle$.
The probability density function $P_F(t)$ of the first-passage time can then be
written as
\begin{align}\label{PDF}
    P_F(t) = -\left\langle 1 \middle\vert \frac{dP^*(t)}{dt} \right\rangle.
\end{align}
The expectation value $E[\tau]$ of the first-passage time is computed as
\begin{align}
    E[\tau] &= \int_{0}^{\infty} t P_F(t) dt \nonumber\\
    &= \int_{0}^{\infty} - t \left\langle 1 \middle\vert \frac{dP^*(t)}{dt} \right\rangle dt \nonumber\\
    &= \int_{0}^{\infty} \langle 1 \vert P^*(t)\rangle dt \nonumber \\
    &= - \langle 1 \rvert (M^*)^{-1} \lvert i\rangle.
\end{align}
Here $(M^*)^{-1}$ denotes the inverse matrix of $M^*$.
The key idea is to divide the states into the goal and the rest and to consider the
modified dynamics in which probability is not conserved.

\section{\label{app:scaling_rates}Scaling Relation of Migration and Mutation Rates}

This appendix applies the first-passage-time formulation of Appendix~\ref{app:fpt} to
derive Eqs.~\eqref{eq.rate_def_v} and \eqref{eq.rate_def_mu}.
For simplicity, only migration in real space is considered.
A single individual performs nearest-neighbor transitions at rate $v_k$.
Specifically, the states are $1, 2, \ldots, k, k+1$; the transition rate for
$l \rightarrow l \pm 1$ is $v_k$, and all other rates vanish.
The transition-rate matrix $M$ is therefore
\begin{equation}
    M = -v_k
    \begin{pmatrix}
        1 & -1 & 0 & & \\
        -1 & 2 & -1 & & & \\
        0 & & \ddots & & 0 \\
         & & -1 & 2 & -1 \\
         & & 0 & -1 & 1
    \end{pmatrix}.
\end{equation}

Following Appendix~\ref{app:fpt}, we take $G = \{k+1\}, G^c = \{1, 2, \ldots, k\}$.
The transition-rate matrix of the modified dynamics and its inverse can then be
written as
\begin{equation}
\begin{aligned}
M^* &=-v_k
\begin{pmatrix}
1&-1&0&&\\
-1&2&-1&&\\
0&-1&2&\ddots&\\
&&\ddots&\ddots&-1\\
&&&-1&2
\end{pmatrix},\\[3pt]
(M^*)^{-1} &=-\frac{1}{v_k}
\begin{pmatrix}
k&k-1&\cdots&1\\
k-1&k-1&\cdots&1\\
\vdots&\vdots&\ddots&\vdots\\
1&1&\cdots&1
\end{pmatrix}.
\end{aligned}
\end{equation}
Since the initial probability vector is $\lvert i\rangle = (1, 0, \ldots, 0)^{\mathsf T}$, the expectation value $\tau_k$ of
the first-passage time to the state $k+1$ is computed as
\begin{align}
    \tau_k &= - \langle 1 \rvert (M^*)^{-1} \lvert i\rangle \\
    &= \frac{k(k+1)}{2v_k}.
\end{align}

The rate $v_k$ is determined by requiring $\tau_k$ to be constant for every $k$.
Since we have fixed $v_1 = v$, $\tau_1$ is also fixed, so that this requirement can be
rewritten as
\begin{align}
    \tau_1 = \tau_k.
\end{align}
From the preceding paragraph,
\begin{align}
    \frac{1}{v} = \frac{k(k+1)}{2v_k},
\end{align}
which gives the relation \eqref{eq.rate_def_v}.
Mutation can be treated in the same way.

\section{\label{app:sat}Derivation of Saturation Time}

This appendix derives the saturation time in Eq.~\eqref{T'_sat}.
As in Appendix~\ref{app:scaling_rates}, it suffices to consider the dynamics of the
population size and the first-passage time of its modified dynamics.
Suppose that there is a single bacterium at some site.
The quantity of interest is the time until the population size reaches the saturated population size
$N$ defined in Eq.~\eqref{eq.N} for the first time, in a situation in which the inflow of
individuals from other sites can be neglected.
Let $Q(n,t)$ be the probability that the population size is $n$ at time $t$.
As the initial condition, we impose $Q(1,0) = 1$ and $Q(n, 0) = 0\;(n>1)$.
Define $\mathbf{Q}(t) = (Q(1, t), Q(2, t), \ldots, Q(K, t))$.
The birth and loss rates are
\begin{align}
    \beta_n = r(1 - \frac{n}{K})n, \qquad \alpha_n = (d + 2v_k)n,
\end{align}
and mutation has been neglected because it is small compared with migration.
Writing the tridiagonal generator as $L_K$, the master equation is
\begin{align}
    \frac{d}{dt}\mathbf{Q}(t) &= L_K\mathbf{Q}(t), \label{eq:population_master}\\
    (L_K)_{ij} &=
    \begin{cases}
        -(\beta_i+\alpha_i), & i=j<K,\\
        -\alpha_K, & i=j=K,\\
        \beta_j, & i=j+1,\\
        \alpha_j, & i=j-1,\\
        0, & \text{otherwise}.
    \end{cases}\nonumber
\end{align}
Setting $G = \{N, N+1, \ldots, K-1, K\}$, the master equation of the modified dynamics
is obtained by restricting $L_K$ to the transient states $1,\ldots,N-1$:
\begin{equation}
    \frac{d}{dt}\mathbf{Q}^*(t) = M_{N-1}^* \mathbf{Q}^*(t),
\end{equation}
\begin{equation}
    M_{N-1}^*=
    (L_K)_{\{1,\ldots,N-1\}\times\{1,\ldots,N-1\}}.
\end{equation}

Next we obtain the first-passage time to $G$ using the modified dynamics.
What is needed for the calculation is the $(i, 1)$ component of
$(M^*_{N-1})^{-1}$ arising from the initial condition, so we compute it.
The components $(M_{N-1}^*)^{-1}_{ij}$ of the inverse matrix can be computed as
\begin{align}\label{eq.saturation_solution}
    ((M_{N-1}^*)^{-1})_{i1}
    &=-\frac{1}{\beta_i}\left[1+\sum_{j=i+1}^{N-1}
    \prod_{\ell=i+1}^{j}\frac{\alpha_\ell}{\beta_\ell}\right].
\end{align}
The first term inside the brackets of Eq.~\eqref{eq.saturation_solution} represents the
contribution of paths along which the population size increases without decreasing on
the way.
The term given by the sum, on the other hand, represents the contribution of paths
along which the population size decreases at least once on the way, through death or
migration to another site.
For $i\ll N$ we have $\alpha_i/\beta_i\ll1$, so that this contribution is small.
For $i\sim N$ the ratio $\alpha_i/\beta_i$ approaches 1 and the approximation does not
hold uniformly for every component, but we have checked numerically that the effect on
the evaluation of $T'_{\mathrm{sat}}$ is small.
In what follows we therefore omit the contribution of paths along which the population
size decreases on the way and approximate
\begin{align}
    ((M_{N-1}^*)^{-1})_{i1}
    \simeq -\frac{1}{\beta_i}.
\end{align}

Following the method of Appendix~\ref{app:fpt}, we obtain the first-passage time until
the population size reaches $N$.
\begin{align}
    T'_{\text{sat}} &= -\sum_{i=1}^{N-1} ((M^*_{N-1})^{-1})_{i,1} \nonumber \\
    &\simeq \sum_{i=1}^{N-1}\frac{1}{\beta_i} \nonumber \\
    &= \sum_{i=1}^{N-1}\frac{1}{r}\left(\frac{1}{i} + \frac{1}{K-i}\right).
\end{align}
When $K_0$ is sufficiently large, so are $K$ and $N$; we therefore use
$\sum_{j=1}^{M} 1/j \simeq \ln M + \gamma\;(\text{for large} M)$, where
$\gamma = 0.5772$ is the Euler--Mascheroni constant.
From the above we obtain
\begin{align}
    T'_{\text{sat}} \simeq \frac{1}{r}\left(\ln\frac{NK}{K-N} + \gamma\right).
\end{align}

\section{\label{app:adapt}Derivation of Adaptation Time}

This appendix gives the intermediate calculations leading to
Eqs.~\eqref{FATPDF}, \eqref{eq.adapt}, and \eqref{T_readapt}.
The definitions and assumptions of the local first-passage problem are given in
Sec.~\ref{sec:tadapt}; here we focus on the generating-function calculation and the
evaluation of the resulting integrals.

\subsection{Survival probability and first-passage-time density}

Starting from the master equation~\eqref{eq.master_eq}, we introduce the generating
function
\begin{align}
    G(z,t) = \sum_{n} z^nP(n,t).
\end{align}
Substitution into Eq.~\eqref{eq.master_eq} gives
\begin{align}
    \frac{\partial G}{\partial t}
    &=-\mu_kN(1-z)G \nonumber\\
    &\quad+\left\{[\delta-v_k-(r'-\mu_k)z](1-z)-v_kz\right\}
    \frac{\partial G}{\partial z}.
\end{align}
This first-order linear partial differential equation has the initial condition
$G(z,0)=1$.
Solving it by the method of characteristics and using $S(t)=G(1,t)$ gives the survival
probability in Eq.~\eqref{FATPDF}; differentiating it gives the corresponding
first-passage-time density in the same equation.

\subsection{Evaluation of $T_{\mathrm{adapt}}$}

Writing the survival probability in Eq.~\eqref{FATPDF} as $S(t)=e^{-af(t)}$, the
integral in Eq.~\eqref{Tadapt_int} becomes
\begin{align}\label{S(t)_transform}
    T_{\text{adapt}} &= \int_{0}^{\infty} e^{-af(t)} dt, \\
    f(t) &:= \ln(b\cosh bt + \mu_k\sinh bt) - \mu_k t - \ln b. \nonumber
\end{align}
We have $f(0)=0$.
The first two derivatives at the boundary $t=0$ are
\begin{align}
    \frac{df(t)}{dt}\left|_{t=0}\right. &= \frac{(b^2-\mu_k^2)\tanh bt}{b + \mu_k\tanh bt}\left|_{t=0}\right. =0, \\
    \frac{d^2f(t)}{dt^2}\left|_{t=0}\right. &=  b^2 - \mu_k^2 >0.
\end{align}
Expanding around $t=0$ therefore gives
$f(t) = \frac{1}{2}(b^2-\mu_k^2)t^2 + \mathcal{O}(t^3)$, and for $a\gg 1$ Laplace's
method yields
\begin{align}
    T_{\text{adapt}} &= \int_{0}^{\infty} e^{-af(t)} dt\nonumber\\
    &\simeq \int_{0}^{\infty} \exp\left[-\frac{a}{2}(b^2-\mu_k^2)t^2\right] dt\nonumber\\
    &= \sqrt{\frac{\pi}{2a(b^2-\mu_k^2)}} = \sqrt{\frac{\pi}{2\mu_k v_k N}}.
\end{align}
This is Eq.~\eqref{eq.adapt}.

The path $(1,1)\rightarrow(2,1)\rightarrow(2,2)$ is also conceivable as an adaptation
process.
The expectation value of its first-passage time can be obtained by considering the
master equation in the same way, but it can be shown to be larger than that of the
path $(1,1)\rightarrow(1,2)\rightarrow(2,2)$~\cite{Hermsen2010}.
We therefore neglect its contribution in this paper.

\subsection{Evaluation of $T_{\mathrm{readapt}}$}

Using Eq.~\eqref{Treadapt_int} and the fact that $T_{\text{adapt}}$ is the mean over the
full first-passage-time distribution, we write
\begin{align}
    T_{\text{readapt}} &= \int_{T_{\text{adapt}}}^{\infty} tF(t)dt \nonumber\\
    &= T_{\text{adapt}} - \int_{0}^{T_{\text{adapt}}}tF(t)dt.\nonumber
\end{align}
Define $\widetilde{S}(u):=S(T_{\text{adapt}}u)$.
Changing variables to $t=T_{\text{adapt}}u$ and integrating by parts gives
\begin{align}
    \int_{0}^{T_{\text{adapt}}}tF(t)dt
    &= -T_{\text{adapt}}\int_{0}^{1}u\frac{d\widetilde{S}(u)}{du}du \nonumber\\
    &= -T_{\text{adapt}}\widetilde{S}(1)
    +T_{\text{adapt}}\int_{0}^{1}\widetilde{S}(u)du.
\end{align}
Substituting Eq.~\eqref{eq.adapt} into Eq.~\eqref{FATPDF} gives
\begin{align}
    \widetilde{S}(u)
    = \left[
    \frac{b\exp(\mu_kT_{\text{adapt}}u)}
    {b\cosh(bT_{\text{adapt}}u)
    +\mu_k\sinh(bT_{\text{adapt}}u)}
    \right]^a.
\end{align}
In the large-$N$ limit, $T_{\text{adapt}}$ is of order $N^{-1/2}$.
Using the expansion of $f(t)$ above together with Eq.~\eqref{eq.adapt}, we find
\begin{align}
    af(T_{\text{adapt}}u) \simeq \frac{\pi}{4}u^2.
\end{align}
Hence
\begin{align}
    \widetilde{S}(u) \simeq e^{-\pi u^2/4}.
\end{align}
Setting
\begin{align}
    C := 1+e^{-\pi/4} - \int_{0}^{1}e^{-\pi u^2/4}du,
\end{align}
we obtain Eq.~\eqref{T_readapt}.

\section{\label{app:scaling_kstar}Derivation of Scaling}

This appendix derives Eq.~\eqref{eq.scaling}.
Differentiating Eq.~\eqref{eq.FATfunc} directly with respect to $k$ and rearranging
under the conditions $vk^2 \ll d$ and $d \ll r$, we obtain
\begin{align}
    \frac{d\mathrm{FAT}(k)}{dk} \simeq -\frac{1}{2}\sqrt{\frac{2\pi}{\mu v K_0}}\frac{1}{x^{3/2}} + \frac{1}{r}\ln\frac{K_0 r}{dx} + \frac{2}{rx} - \frac{1}{r},
\end{align}
where $x = k + 1$.
Define
\begin{align}
    H(x) := -\frac{1}{2}\sqrt{\frac{2\pi}{\mu v K_0}}\frac{1}{x^{3/2}} + \frac{1}{r}\ln\frac{K_0 r}{dx} + \frac{2}{rx} - \frac{1}{r}.
\end{align}

By the implicit function theorem,
\begin{align}
    \frac{\partial \ln x}{\partial \ln p} = -\frac{pH_p}{xH_x}.
\end{align}
Computing this for $p = r, v, \mu, d, K_0$ gives
\begin{align}\label{implicit_result}
    \frac{\partial \ln x}{\partial \ln p}=
    \begin{cases}
        \frac{2\Lambda -4 +4/x}{3\Lambda- 5 + 2/x}\;\;(p=r)\\
        -\frac{\Lambda -1 + 2/x}{3\Lambda - 5 + 2/x}\;\;(p=\mu, v)\\
        -\frac{\Lambda + 1 + 2/x}{3\Lambda -5 +2/x}\;\;(p=K_0)\\
        \frac{2}{3\Lambda - 5 + 2/x}\;\;(p=d)
    \end{cases},
\end{align}
where $\Lambda = \ln (K_0r/dx)$ and the stationarity condition $H(x) = 0$ has been
used in rearranging.
Here $1 < x = k^* + 1 \ll k_{\text{max}} = \sqrt{2d/v} \ll K_0$.
Assuming $x\ll K_0$ and $2/x \leq \Lambda$, Eq.~\eqref{implicit_result} can be written as
\begin{align}
    \frac{\partial \ln x}{\partial \ln p}=
    \begin{cases}
        \frac{2}{3}\;\;(p=r)\\
        -\frac{1}{3}\;\;(p=\mu, v)\\
        -\frac{1}{3}\;\;(p=K_0)\\
        0\;\;(p=d)
    \end{cases}.
\end{align}
The scaling is therefore
\begin{align}
    k^* + 1 \propto \left(\frac{r^2}{\mu v K_0}\right)^{\frac{1}{3}},
\end{align}
which is also consistent with $x \ll K_0$.

\section{\label{app:geomean}Origin of the Geometric Mean}

This appendix uses toy models to explain how population effects generate a geometric
mean of rates, such as $\sqrt{\mu v}$.
To contrast the two mechanisms, we consider an OR circuit and an AND circuit.
The OR circuit has a start state and two goal states, labeled 1 and 2.
A particle makes stochastic transitions among these three states.
The rate of $\text{start}\rightarrow\text{goal 1}$ is $\mu$ and the rate of
$\text{start}\rightarrow\text{goal 2}$ is $v$.
In this circuit a particle reaches either goal 1 or goal 2 from the start state
through a stochastic process.
For the OR circuit we consider the first-passage time until either goal is reached.
The AND circuit has a start state, an intermediate state, and a goal state.
Again, a particle makes stochastic transitions among these three states.
The rate of $\text{start}\rightarrow\text{intermediate}$ is $\mu$ and the rate of
$\text{intermediate}\rightarrow\text{goal}$ is $v$.
In this circuit a particle reaches the intermediate state from the start state through a
stochastic process and then reaches the goal.
For the AND circuit we consider the first-passage time until the goal is reached.

First, consider the statistics of the first-passage time for the OR circuit, specifically
the time until one of $N$ particles reaches goal 1 or goal 2.
Let $f_{\text{OR}}(t)$ be the probability density function of the FPT for $N=1$.
The probability that a goal has been reached by time $t$ is its cumulative distribution
function $\mathrm{Cum}_{\text{OR}}(t)$.
\begin{align}
    \mathrm{Cum}_{\text{OR}}(t) = \int_{0}^{t} f_{\text{OR}}(s) ds.
\end{align}
The probability $R_{\text{OR}}(t)$ that none of the $N$ particles has reached goal 1 or
goal 2 can therefore be written as
\begin{align}
    R_{\text{OR}}(t) = \left(1 - \mathrm{Cum}_{\text{OR}}(t)\right)^N.
\end{align}
The probability $P_s(t)$ that a particle is in the start state is
\begin{align}
    P_s(t) = e^{-(\mu + v)t},
\end{align}
and therefore the first-passage-time density is
\begin{align}
    f_{\text{OR}}(t) = (\mu + v)e^{-(\mu + v)t}.
\end{align}

The corresponding quantities are next considered for the AND circuit.
Let $f_{\text{AND}}(t)$ be the probability density function of the FPT for $N=1$,
$\mathrm{Cum}_{\text{AND}}(t)$ its cumulative distribution function, and
$R_{\text{AND}}(t)$ the probability that none of the $N$ particles has reached the
goal.
Explicitly,
\begin{align}
    f_{\text{AND}}(t) &= \frac{\mu v}{v - \mu}(e^{-\mu t} - e^{-vt}) \\
    \mathrm{Cum}_{\text{AND}}(t) &= \int_{0}^{t} f_{\text{AND}}(s) ds \\
    R_{\text{AND}}(t) &= \left(1 - \mathrm{Cum}_{\text{AND}}(t)\right)^N.
\end{align}

For a single particle, the mean first-passage times
$\text{FPT}_{\text{OR}, 1}$ and $\text{FPT}_{\text{AND}, 1}$ can be computed as
follows.
\begin{align}
    \text{FPT}_{\text{OR}, 1} &= \frac{1}{\mu + v}\\
    \text{FPT}_{\text{AND}, 1} &= \frac{1}{\mu} + \frac{1}{v}.
\end{align}
Next we consider $N$ particles; for sufficiently large $N$,
\begin{align}
    R_{\text{OR}}(t) &\simeq e^{-N\mathrm{Cum}_{\text{OR}}(t)}\\
    R_{\text{AND}}(t) &\simeq e^{-N\mathrm{Cum}_{\text{AND}}(t)}.
\end{align}
Expanding these around $t=0$ gives
\begin{align}
    R_{\text{OR}}(t) &\simeq e^{-N(\mu + v)t}\\
    R_{\text{AND}}(t) &\simeq e^{-N\frac{\mu v}{2}t^2}.
\end{align}
The respective mean first-passage times in the limit of large $N$ are therefore
\begin{align} 
    \text{FPT}_{\text{OR}, N} &= \frac{1}{N(\mu + v)}\\
    \text{FPT}_{\text{AND}, N} &= \sqrt{\frac{\pi}{2N \mu v}}.
\end{align}
From the above, the geometric mean appearing in Eqs.~\eqref{cond_min} and
\eqref{eq.scaling} can be attributed both to the sequential process of mutation
followed by migration and to the population effect.


\begin{thebibliography}{99}
 
\bibitem{AMRCollaborators2022}
Antimicrobial Resistance Collaborators,
Global burden of bacterial antimicrobial resistance in 2019: a systematic analysis,
\textit{Lancet} \textbf{399}, 629 (2022).

\bibitem{Zhang2011}
Q.~Zhang, G.~Lambert, D.~Liao, H.~Kim, K.~Robin, C.-K.~Tung, N.~Pourmand, and R.~H.~Austin,
Acceleration of emergence of bacterial antibiotic resistance in connected microenvironments,
\textit{Science} \textbf{333}, 1764 (2011).
 
\bibitem{Baym2016}
M.~Baym, T.~D.~Lieberman, E.~D.~Kelsic, R.~Chait, R.~Gross, I.~Yelin, and R.~Kishony,
Spatiotemporal microbial evolution on antibiotic landscapes,
\textit{Science} \textbf{353}, 1147 (2016).
 
\bibitem{Leale2018}
A.~M.~Leale and R.~Kassen,
The emergence, maintenance, and demise of diversity in a spatially variable antibiotic regime,
\textit{Evol.~Lett.} \textbf{2}, 134 (2018).
 
\bibitem{Hermsen2010}
R.~Hermsen and T.~Hwa,
Sources and sinks: a stochastic model of evolution in heterogeneous environments,
\textit{Phys.\ Rev.\ Lett.} \textbf{105}, 248104 (2010).
 
\bibitem{Hermsen2012}
R.~Hermsen, J.~B.~Deris, and T.~Hwa,
On the rapidity of antibiotic resistance evolution facilitated by a concentration gradient,
\textit{Proc.\ Natl.\ Acad.\ Sci.\ U.S.A.} \textbf{109}, 10775 (2012).
 
\bibitem{Hermsen2016}
R.~Hermsen,
The adaptation rate of a quantitative trait in an environmental gradient,
\textit{Phys.\ Biol.} \textbf{13}, 065003 (2016).
 
\bibitem{Greulich2012}
P.~Greulich, B.~Wac{\l}aw, and R.~J.~Allen,
Mutational pathway determines whether drug gradients accelerate evolution of drug-resistant cells,
\textit{Phys.\ Rev.\ Lett.} \textbf{109}, 088101 (2012).
 
\bibitem{Piskovsky2023}
V.~Piskovsky and N.~M.~Oliveira,
Bacterial motility can govern the dynamics of antibiotic resistance evolution,
\textit{Nat.\ Commun.} \textbf{14}, 5584 (2023).
 
\bibitem{Gillespie1983}
J.~H.~Gillespie,
Some properties of finite populations experiencing strong selection and weak mutation,
\textit{Am.\ Nat.} \textbf{121}, 691 (1983).
 
\bibitem{Gillespie1984}
J.~H.~Gillespie,
Molecular evolution over the mutational landscape,
\textit{Evolution} \textbf{38}, 1116 (1984).
 
\bibitem{Atwood1951}
K.~C.~Atwood, L.~K.~Schneider, and F.~J.~Ryan,
Periodic selection in \textit{Escherichia coli},
\textit{Proc.\ Natl.\ Acad.\ Sci.\ U.S.A.} \textbf{37}, 146 (1951).
 
\bibitem{Sniegowski2010}
P.~D.~Sniegowski and P.~J.~Gerrish,
Beneficial mutations and the dynamics of adaptation in asexual populations,
\textit{Philos.\ Trans.\ R.\ Soc.\ B} \textbf{365}, 1255 (2010).
 
\bibitem{Desai2007}
M.~M.~Desai and D.~S.~Fisher,
Beneficial mutation--selection balance and the effect of linkage on positive selection,
\textit{Genetics} \textbf{176}, 1759 (2007).
 
\bibitem{Sekimoto2022}
K.~Sekimoto,
Derivation of the first passage time distribution for Markovian process on discrete network,
arXiv:2110.02216.

\bibitem{Sezonov2007}
G.~Sezonov, D.~Joseleau-Petit, and R.~D'Ari,
\textit{Escherichia coli} physiology in Luria--Bertani broth,
\textit{J.\ Bacteriol.} \textbf{189}, 8746 (2007).

\bibitem{Kurzthaler2024}
C.~Kurzthaler et al.,
Characterization and control of the run-and-tumble dynamics of
\textit{Escherichia coli},
\textit{Phys.\ Rev.\ Lett.} \textbf{132}, 038302 (2024).

\bibitem{Lee2012}
H.~Lee, E.~Popodi, H.~Tang, and P.~L.~Foster,
Rate and molecular spectrum of spontaneous mutations in the bacterium
\textit{Escherichia coli} as determined by whole-genome sequencing,
\textit{Proc.\ Natl.\ Acad.\ Sci.\ U.S.A.} \textbf{109},
E2774 (2012).

\bibitem{Regoes2004}
R.~R.~Regoes, C.~Wiuff, R.~M.~Zappala, K.~N.~Garner,
F.~Baquero, and B.~R.~Levin,
Pharmacodynamic functions: a multiparameter approach to the design
of antibiotic treatment regimens,
\textit{Antimicrob.\ Agents Chemother.} \textbf{48},
3670 (2004).
 
\bibitem{Sandegren2008}
L.~Sandegren, A.~Lindqvist, G.~Kahlmeter, and D.~I.~Andersson,
Nitrofurantoin resistance mechanism and fitness cost in
\textit{Escherichia coli},
\textit{J.\ Antimicrob.\ Chemother.} \textbf{62}, 495 (2008).

\bibitem{Andersson2010}
D.~I.~Andersson and D.~Hughes,
Antibiotic resistance and its cost: is it possible to reverse resistance?,
\textit{Nat.\ Rev.\ Microbiol.} \textbf{8}, 260--271 (2010).

\bibitem{Schrag1997}
S.~J.~Schrag, V.~Perrot, and B.~R.~Levin,
Adaptation to the fitness costs of antibiotic resistance in
\textit{Escherichia coli},
\textit{Proc.\ R.\ Soc.\ Lond.\ B} \textbf{264}, 1287 (1997).
 
\end{thebibliography}
\end{document}